\documentclass[reprint,prl,preprintnumbers,superscriptaddress,longbibliography,nofootinbib]{revtex4-2}
\usepackage{orcidlink}
\usepackage{graphicx}% Include figure files
\usepackage{dcolumn}% Align table columns on decimal point
\usepackage{bm}% bold math
\usepackage{textgreek}
\usepackage{placeins}
\usepackage{braket}
\usepackage{physics}
\usepackage{overpic}
\usepackage{amssymb}
\usepackage{amsmath}
\usepackage{subfigure}
\usepackage{qcircuit}
\usepackage{tikz}
\usepackage{adjustbox}
\usepackage{longtable}
\usepackage{soul}
\usepackage{amssymb}
\usepackage{cleveref}

\begin{document}

\title{Efficient use of quantum computers for collider physics}
\author{Christian W.~Bauer \,\orcidlink{0000-0001-9820-5810}}
\email{cwbauer@lbl.gov}
\affiliation{Physics Division, Lawrence Berkeley National Laboratory, Berkeley, California 94720, USA}
\affiliation{Department of Physics, University of California, Berkeley, Berkeley, CA 94720}
\date{\today}
\begin{abstract}
Most observables at particle colliders involve physics at a wide variety of distance scales. 
Due to asymptotic freedom of the strong interaction, the physics at short distances can be calculated reliably using perturbative techniques, while long distance physics is non-perturbative in nature. 
Factorization theorems separate the contributions from different scales scales, allowing to identify the pieces that can be determined perturbatively from those that require non-perturbative information, and if the non-perturbative pieces can be reliably determined, one can use experimental measurements to extract the short distance effects, sensitive to possible new physics.
Without the ability to compute the non-perturbative ingredients from first principles one typically identifies observables for which the non-perturbative information is universal in the sense that it can be extracted from some experimental observables and then used to predict other observables. 
In this paper we argue that the future ability to use quantum computers to calculate non-perturbative matrix elements from first principles will allow to make predictions for observables with non-universal non-perturbative long-distance physics.

\end{abstract}
\maketitle

High energy physics (HEP) is concerned with testing physics at short distances and and therefore high energies. 
A commonly used experimental tool to test the most fundamental constituents of nature and the forces among them is to collide particles at high energies, allowing insights into the short distance interactions.
Describing high-energy particle collisions requires understanding the effects originating from a wide variety of energy scales. 
For most observables, collisions at the LHC require to understand interactions ranging from the largest scale in the problem, namely the energy of the colliding protons $E = 7 \, {\rm TeV}$ all the way to the lowest hadronic scales of order the mass mass of the pion $E \sim 100 \, {\rm MeV}$ or even lower.
Asymptotic freedom implies that the strong coupling is small at large scales and large at small scales, with the transition point being the scale $\Lambda_{\rm QCD} \sim {\cal O}(100)$ MeV.
The short distance physics HEP is interested in testing can therefore be calculated reliably using perturbation theory. 
However, as mentioned, predictions of any relevant observable depend not only on the short distance, but also on long distance physics, and in particular those at or below $\Lambda_{\rm QCD}$ require a non-perturbative treatment.
Being able to predict these long distance effects is therefore of parmount importance to test short distance interactions and possibly discover deviations from the Standard Model.

The most common way to make predictions for collider observables is to use parton shower algorithms~\cite{Sjostrand:2006za,Bahr:2008pv,Gleisberg:2008ta}, which combine a perturbative treatment of short distance physics with a probabilistic parton shower algorithm describing emissions at smaller, but still perturbative energies and a  model~\cite{Andersson:1983jt,Webber:1983if} describing the long distance hadronization process.
While the original parton shower algorithms included only perturbative information at the lowest order, there has been tremendous progress to combine higher perturbative fixed-order and resummed calculations  with parton showers, and even combining high-order resummed calculations.
The non-perturbative model contains a number of free parameters which are tuned to reproduce a wide set of experimental data.

A more systematic approach go get control over long distance physics that works for many observables is the use of factorization theorems, which separate the short- and long-distance physics systematically. 
The derivation of factorization theorems has a long history~\cite{Collins:1981uk,Collins:1981ta,Collins:1987pm,Collins:1989gx}, and a modern way of deriving factorization theorems~\cite{Bauer:2002nz,Bauer:2008jx,Lee:2006fn,Bauer:2008dt,Jouttenus:2011wh} is using effective field theories, in particular Soft-Collinear Effective Theory (SCET)~\cite{Bauer:2000ew,Bauer:2000yr,Bauer:2001ct,Bauer:2001yt} for the case of collider observables. 
The important aspect of factorization theorems is that it allows to characterize long-distance physics by matrix element of well-defined operators defined in SCET, and typically the number of such matrix elements is manageable for typical observables of interest.
Observables are therefore predicted in terms of long distance matrix elements, combined with the short distance physics one is after. 
If these long distance matrix elements can be reliably determined, it allows to use the experimental data to extract the short distance physics one is after.

Traditionally, factorization theorems have proven most useful when the non-perturbative ingredients are universal, in the sense that the same non-perturbative matrix element contributes to more than one experimentally accessible process. 
Besides hadron masses, the most well known universal non-perturbative objects are parton distribution functions, which describe non-perturbative effects from collinear radiation. 
Since parton distribution functions contribute to almost all observables at hadron colliders, they are extracted from a measurement of a range of observables and then used as inputs for other measurements of interest. 
A second example are so-called shape functions, which describe non-perturbative effects from soft radiation. 
For example, in the decays of $B$ mesons to light particles, a single shape function describes both the shape of the photon spectrum in inclusive $B \to X_s \gamma$ decays, as well as at the endpoint of the lepton spectrum in $B \to X_u \ell \nu$ decays~\cite{Neubert:1993ch,Neubert:1993um,Bauer:2003pi,Ligeti:2008ac}.

However, many other observables of interest depend on non-perturbative information that is specific to the given observable. As an example, consider one of the simplest observables in collider physics, namely event shape distributions at lepton colliders~\cite{Brandt:1964sa,Farhi:1977sg,Catani:1992jc,Catani:1991bd,Clavelli:1981yh,Ellis:1980wv}. 
Such event shape distributions are typically obtained by measuring the momenta of all final state particles in an $e^+ e^-$ collision, and then computing a certain function of these momenta. 
The most common event shape is thrust~\cite{Brandt:1964sa,Farhi:1977sg}, obtained by summing the inner product of all momenta $p_i$ with a fixed axis $\hat t$, dividing by the sum of the absolute values of these inner products. 
One then finds the axis that maximizes the expression (resulting in the thrust axis) and this maximized result is value of the thrust event shape. 
This can be written as
\begin{align}
    T = {\rm max}_{\hat t}\frac{\sum_i \hat t \cdot p_i}{\sum_i |\hat t \cdot p_i|}
    \,.
\end{align}
Another commonly used event shape is the $C$ parameter, defined as~\cite{Ellis:1980wv}
\begin{align}
    C = \frac{3}{2} \frac{\sum_{ij} |p_i| |p_j| \sin^2\theta_{ij}}{(\sum_i |p_i|)^2}
    \,,
\end{align}
where $\theta_{ij}$ is the angle between the momenta $p_i$ and $p_j$.
Both of these event shapes approach 1 in the limit where all final state particles are contained in two back-to-back jets. 
In this limit one can derive the factorization theorem
\begin{align}
\label{eq:ThrustFactorization}
    \frac{{\rm d} \sigma}{{\rm d}e} = \int\! {\rm d}k \, \frac{{\rm d}\sigma_{\rm pert}}{{\rm d} e} \left(e - \frac{k}{Q} \right)F(k) + {\cal O}\left( \frac{k}{Q} \right)
    \,,
\end{align}
where $e \equiv 1-T$ or $e \equiv 1-C$, depending on which event shape is being studied.
This factorization states that the differential thrust distribution is obtained through a convolution of a perturbatively calculable cross section ${\rm d} \sigma_{\rm pert} / {\rm d} e$, which describes the short distance physics above the soft scale
\begin{align}
    \Lambda_s \sim Q \, e\,,
\end{align}
and a non-perturbative object called a soft- or shape-function $F(k)$ describing the physics below that scale. 
As discussed before, in order for the event shape observables to be able to test the short-distance physics in ${\rm d}\sigma_{\rm pert} / {\rm d} e$ requires knowledge of the soft function $F(k)$. 

One way to obtain information about the non-perturbative information $F(k)$ is to measure it experimentally.
This is possible if the same function $F(k)$ contributes to more than one observable. Such a universal soft function $F(k)$ can then be extracted by measuring one observable, and can then be used to predict another observable. 
While the first moment of several shape-functions can be shown to be universal, the full non-perturbative shape function depends on the event-shape being considered.
For more complicated observables, the non-perturbative information required will be contained in more complicated soft and/or collinear functions, which are typically not universal.
Making first principles predictions of the corresponding observables is therefore very difficult or even impossible.

An alternative to determining non-perturbative information from data is to compute it numerically.
The only known technique to compute non-perturbative quantities from first principles in QCD is lattice QCD (LQCD), which uses the path integral to calculated observables in a discretized space-time. 
Since the dimensionality of the integral is exponential in the number of lattice sites, this path integral is only possible to perform using Monte-Carlo integration techniques. 
This requires to perform calculations in Euclidean space, in order to render the integrand positive definite and avoid a sign problem~\cite{Troyer:2004ge,Alexandru:2016gsd}.
This, however, makes the calculation of real-time dynamics and observables that are inherently Minkowskian in nature inaccessible to traditional LQCD methods.
It should be noted that there has been significant progress towards relating parton distribution functions to the expectation value of purely spatial correlators using the Large Momentum Effective Theory~\cite{Ji:2013dva,Ji:2014gla,Ma:2014jla,Orginos:2017kos,Ji:2020ect}. 
However, this approach is not applicable to soft functions discussed in this work, since they involve multiple light-like directions. For this reason, computing the non-perturbative information of event shapes and related processes is not possible using known classical numerical techniques.

In seminal work by Jordan, Lee and Preskill~\cite{Jordan:2012xnu,Jordan:2011ci} it was shown that quantum algorithms are able to simulate the dynamics of quantum field theories efficiently, with resources that scale polynomial in the number of lattice sites. 
This has resulted in a tremendous amount of research on Hamiltonian Lattice Gauge Theories (HLGT)s~\cite{Kogut1975,Byrnes:2005qx}, ranging from ${\rm U}(1)$~\cite{Banerjee:2012pg,Hauke:2013jga,Zohar:2013zla,Kuhn:2014rha,Kasper:2015cca,Zohar:2015hwa,Martinez:2016yna,Yang:2016hjn,Kokail:2018eiw,Klco:2018kyo,Lu:2018pjk,Kaplan:2018vnj,Mil:2019pbt,Davoudi:2019bhy,Surace:2019dtp,Haase:2020kaj,Luo:2019vmi,Shaw:2020udc,Yang:2020yer,Ott:2020ycj,Paulson:2020zjd,Nguyen:2021hyk,Zhou:2021kdl,Riechert:2021ink,Bauer:2021gek,Kane:2022ejm,Grabowska:2022uos,zhang2023observation,Farrell:2023fgd,Nagano:2023uaq}, ${\rm SU}(2)$~\cite{Zohar:2012xf,Stannigel:2013zka,Mezzacapo:2015bra,Mathur:2015wba,Raychowdhury:2018osk,Raychowdhury:2019iki,Klco:2019evd,Dasgupta:2020itb,Davoudi:2020yln,Atas:2021ext,ARahman:2021ktn,Osborne:2022jxq,Davoudi:2022xmb,halimeh2022gauge,ARahman:2022tkr,zache2023quantum,Alexandru:2023qzd,DAndrea:2023qnr,Turro:2024pxu} to ${\rm SU}(3)$~\cite{Anishetty:2009nh,Alexandru:2019nsa,Kan:2021xfc,Ciavarella:2021nmj,Farrell:2022wyt,Farrell:2022vyh,Atas:2022dqm,Ciavarella:2021lel,Ciavarella:2023mfc,hayata2023qdeformedformulationhamiltonian,Farrell:2024fit,Ciavarella:2024fzw,Balaji:2025afl,Ciavarella:2025bsg}.
The expectation of the field is that with significant progress on the theoretical development of HLGTs, the corresponding quantum algorithms~\cite{Hariprakash:2023tla,Rhodes:2024zbr,Spagnoli:2024mib,Kane:2024odt,Gomes:2024tup,Hardy:2024ric} as well as quantum hardware, first principle calculations of non-perturbative QCD dynamics will become possible. 

In principle, it is possible to directly calculate the full $S$-matrix corresponding to the collider under consideration. 
The basic mechanism is to prepare the initial state of the collisions (typically two wave packets moving towards each other), evolve this system forward in time using the Hamiltonian of the Standard Model, and then performing a measurement for the desired observable. 
As long as the evolution is applied for a sufficiently long time, such that the initial state is in the ``infinite'' past and the final state in the ``infinite'' future, the frequency of measuring a given value of the final state observable corresponds to the differential cross section.
All steps in this calculation should in principle be calculable efficiently on a quantum computer, as was shown explicitly for a scalar field theory in~\cite{Jordan:2012xnu,Jordan:2011ci}.

The main problem is that such a simulation requires way too many lattice sites to be feasible on any realistic time scale.
This can easily be understood by the fact that the lattice spacing required has to be inversely proportional to the largest energy scale one aims to capture, while the size of the lattice has to be inversely proportional to the smallest scale. 
This implies that the total number of lattice sites is given by
\begin{align}
\label{eq:nL}
    n_L \sim \left( \frac{E_{\rm max}}{E_{\rm min}}\right)^3
    \,.
\end{align}
In order to simulate the entirety of a collision at a collider requires simulating all scales ranging from the lightest hadron and below $E_{\rm min} \sim 100 {\rm MeV}$ to the collider energy.
This requires an immense number of lattice points, ranging from $n_L \gtrsim 10^9$ for LEP and $n_L \gtrsim 10^{15}$ for the LHC. 
Given that the number of logical qubits is likely at least an order of magnitude larger than the number of lattice sites, it is clear that such complete simulations are out of reach in any foreseeable future.

While it might be disappointing that full quantum simulations of collider events seem out of reach, one should question the usage of quantum computers to simulate the physics at short distance scales far above $\Lambda_{\rm QCD}$. At such scales, perturbative calculations are clearly possible and the relevant physics can be calculated with sufficient precision, certainly on a time-scale where quantum computers become available. 
Given that the available quantum computer hardware will likely remain precious for a long time, one should aim to use quantum computers only for those effects that are inaccessible to traditional methods.
We will now argue that in the context of collider physics quantum computers are likely best used to calculate the non-perturbative matrix elements arising in factorization theorems such as~\cref{eq:ThrustFactorization}.

As already discussed, the non-perturbative ingredient of event shape distributions and many other observables are contained in the soft function $F_e(e_s)$, which depends on the details of the observable being measured.
The precise field theoretical definition of the soft function is given as the matrix element of an operator containing soft Wilson lines 
\begin{align}
\label{eq:softfunction}
    F_e(e_s) & = |\langle X(e_s) | T[Y_{n}^\dagger Y_{\bar n}] | 0 \rangle|^2
    \,,
\end{align}
where $X_s$ denotes a given final state which has event shape equal to $e_s$.
The unitary Wilson line operator
\begin{align}
    Y_n = {\rm P} \exp\left[ i g \int_0^\infty \!\!\! {\rm d} s \, n\cdot A(n s) \right]
    \,,
\end{align}
is given in terms of the path ordered exponential of integral over the field operator $A$ of the gluon field along some light-like trajectory $n$.
Since this soft function is the absolute value square of an amplitude, it is positive definite and normalized to unity
\begin{align}
    \int_0^1 \!\! {\rm d} e_s F_e(e_s) = 1
    \,.
\end{align}
In other words, $F_e(e_s)$ is given by a non-perturbative probability distribution. 
One can use a quantum computer to directly sample from this probability distribution, by using three steps.
While the details of such a procedure have only been worked out for a scalar field theory~\cite{Bauer:2021gup}, we will discuss the basic steps required next.

One first creates the ground state of the interacting QCD vacuum $\ket{0}$. 
While this is not a simple procedure, and the details of how to achieve this most efficiently have not been worked out, it has been shown that obtaining the ground state of an interacting scalar field theory can be obtained using polynomial resources, and it is expected that this resource scaling remains true for gauge theories such as QCD as well. 
Possible algorithms include preparing a simpler ground state and adiabatically transitioning to the ground state of QCD, or by setting up a variational algorithm. 
In a second step, one applies the unitary Wilson line operator $T[Y_{n}^\dagger Y_{\bar n}]$.
In a final step one measures the event shape of the final state.
This will collapse the wave function, and the frequency with which a given value of the event shape $e_s$ is observed will be equal to the desired probability distribution.
We will discuss some ideas of performing such a measurement later.

An important observation is that the soft function defined in~\cref{eq:softfunction} describes the physics below the scale $\Lambda_s = Q \, e$, which implies that the lattice spacing required to perform the above steps needs to be $a \sim 1 / \Lambda_s$, and using~\cref{eq:nL} the number of lattice sites is not parametrically enhanced as for the case of computing the full scattering process. 
For different values of $e_s$ one can further factorize the soft function into a non-perturbative and perturbative contribution. 
This implies that calculating the soft function, and in fact, as explained later, directly sampling from the soft function will be possible on quantum computers with many orders of magnitude less resources that it would require to compute the full cross section.

The main ingredient in the discussion so far has been the presence of a factorization theorem for the event shape observable. 
In~\cref{eq:ThrustFactorization} we gave the factorization theorem as a convolution over a perturbatively calculable term and a non-perturbative shape function, but the general factorization theorem is more broad and separates out the physics from hard, collinear and soft physics. For processes with two jets in the final state this can schematically be written as
\begin{align}
    \frac{{\rm d} \sigma}{{\rm d}e} = \left[h \otimes J_1(e_1) \otimes J_2(e_2) \otimes F_e(e_s) \right](e)
    \,,
\end{align}
and processes with more jets this can be generalized by including more jet functions.
The $\otimes$ denotes a convolution over the various functions, each of which are evaluated at their own value $e_i$. The functions are combined in such a way that the final expression corresponds to the value of the observable, which is denoted by the notation $(e)$. The details of these functions is not too important for this discussion.
The hard and jet functions are perturbatively calculable, which is why they were combined in~\cref{eq:ThrustFactorization}.

While such a factorization theorem takes a particularly simple form for event shapes at lepton colliders, factorization theorems similar to the one above are possible for a large variety of observables. 
For observables at hadron colliders, additional non-perturbative functions, namely parton distribution functions, are required, but as discussed earlier, these can be extracted much easier using experimental data or directly computed using classical computers. 
The main requirement for such a factorization theorem to hold is that one considers jet observables, for which hadronic final states are dominated by energetic particles moving at small angles relative to one another (collinear) or particles with small energy (soft). For the case of event shapes, this is the case for $e \to 1$ where the factorization theorem holds, since the event becomes 2-jetlike in this region.

A generic formalism for obtaining factorization theorems for jet observables was developed in~\cite{Bauer:2008jx}.
The basic idea is to define a general, fully differential cross energy distribution
\begin{align}
    \frac{{\rm d}\sigma}{{\rm d}\omega} = \frac{1}{2P_I^2} \sum_X \bra{X} {\cal O} \ket{I} \delta\left[\omega - \omega(X)\right]
    \,.
\end{align}
Here $I$ and $X$ denote the initial and final states with momenta $p_I$ and $p_X$, ${\cal O}$ denotes the operator mediating the underlying short distance interaction, while $\omega$ represents the energy distribution of all final state particles in the event.
Considering only massless particles, most observables can be constructed from knowledge of the the fully differential energy distribution
\begin{align}
    \frac{{\rm d}\sigma}{{\rm d}e} = \int \! {\rm d} \omega \, \frac{{\rm d}\sigma}{{\rm d}\omega} \, \delta(e - e[\omega]) 
    \,.
\end{align}
While we write this expression for an event shape observable $e$, it holds for more general observables as well, and for the remainder of this paper we will write expressions for such general observables $O$. 

Using steps very similar to the proof of factorization for event shape distributions, one can obtain a general factorization theorem for this fully differential energy distribution, valid for final state configurations that are jet-like
\begin{align}
    \frac{{\rm d} \sigma}{{\rm d}\omega} = \left[h \otimes J_1(\omega_1) \cdots\otimes J_n(\omega_2) \otimes F_\omega(\omega_s) \right](\omega)
    \,.
\end{align}
Each function is evaluated at its own energy distribution $\omega_i$ and, as before, the convolution is defined such that the combination corresponds to the desired energy distribution, as indicated by the $(\omega)$.
For lepton colliders the jet functions run only over final state jets, while for hadron colliders initial state jets are also present, containing non-perturbative information through parton distribution functions.

While the factorization theorem for the energy distribution is rather general, factorization breaks for the most general observables since the observable itself can mix collinear and soft contributions in a non-factorizable way. On the other hand, a wide class of observables exist for which such a factorization theorem can be obtained, and for all of these observables the corresponding soft function can be written as
\begin{align}
	F_O[o_s] = \int \! {\rm d} \omega_s \, F_\omega(\omega_s) \delta[o_s - O_s(\omega)]
	\,,
\end{align}
where the function $O(\omega)$ describes how soft radiation is contributing to the observables under consideration. 
The general soft function describing the energy distribution can be written schematically similar to the soft function discussed above
\begin{align}
    F_\omega(\omega_s) & = |\langle X(\omega_s) | T[Y_{n}^\dagger Y_{\bar n}] | 0 \rangle|^2
    \,,
\end{align}
where now $\ket{X(\omega_s)}$ denotes a final state with a particular energy distribution. 
For a particular observable $O$, the corresponding soft function becomes
\begin{align}
    F_O(o_s) & = |\langle X(o_s) | T[Y_{n}^\dagger Y_{\bar n}] | 0 \rangle|^2
    \,.
\end{align}
A quantum computer can therefore sample from the fully exclusive soft function by directly measuring the full energy of the final state, or from the soft function of any observable that can be constructed from these momenta. 
Note that this is at least in principle independent of the details of the soft function, and in particular does not require the soft function to be universal.
This implies that it is useful to carefully define factorization theorems for jet observables which are sensitive to the underlying physics one is trying to explore, rather than worrying about whether the non-perturbative physics, in particular that contained in the soft function, will spoil the predictability of a measurement.

We end by discussing how to design optimal quantum algorithms and hardware to compute the soft functions discussed in this work.
The energy distribution defining the fully differential soft function is defined through an energy flow operator acting on the final state~\cite{Lee:2006nr}
\begin{align}
	{\cal E}(\hat n) \ket{X} = \sum_i \, \omega_i \, \delta(\hat n - \hat n_i) \ket{X}
	\,.
\end{align}
This energy flow operator can be related to the energy momentum tensor as~\cite{Sveshnikov:1995vi,Korchemsky:1997sy}
\begin{align}
	{\cal E}(\hat n) = \lim_{R \to \infty} R^2 \int_0^\infty\!\!\! {\rm d} t \, \hat n_i T_{0i}(t, R \hat n)
	\,.
\end{align}
This operator therefore measures the total energy in radial direction $\hat n$ an infinite distance from the origin, integrated over all time. 
In a lattice simulation, infinite distance corresponds to the boundary of the lattice, such that one needs to measure the distribution of energy at the boundary of the lattice.
This implies that in a quantum HLGT simulation one needs to continuously read out the energy of the qubits describing the fields at the edge of the lattice, extract the energy and integrate this over the simulation time. 
A calculation of the soft function therefore requires measuring qubits only at the edge of the lattice, while the interior of the lattice only serves to describes how the Wilson line operators act on the vacuum initial state, giving rise to energy flowing towards the boundary.
Since the energy flow operator is defined through the energy flow at the boundary a distance $R \to \infty$ away from the collision point, there will be finite volume effects arising from finite size of the lattice. 
To mitigate these effects, one has to choose the lattice large enough for this distance to be larger than the the typical QCD distance scale $1 / \Lambda_{\rm QCD}$. 

In summary, we have put forth a vision for how quantum computers are utilized most efficiently to make predictions for collider physics. 
We started by showing that while in principle the full scattering could be simulated on a quantum computer, the number of lattice sites scales with the ratio of largest to smallest energy scale in the problem, making such a simulation too costly to be possible on any reasonable time scale. 
We then discussed how factorization theorems can be derived using well established principles that separate the different scales in the problem and in particular isolate the soft physics contributing into a non-perturbative probability distribution. 
A quantum computer would be able to sample from this probability distribution, which combined with well-defined perturbative calculations will lead to a first principles prediction for any observable for which a factorization theorem is available. 
An interesting aspect is that for measurements of this kind, the qubits comprising the lattice used in the quantum simulation would only need to be measured at the boundary of the lattice.

\acknowledgments
We would like to thank Iain Stewart for conversations about Large Momentum Effective Theory, as well as Anthony Ciavarella, Zohreh Davoudi and Zoltan Ligeti for comments on the manuscript. 
This material is based upon work supported by the U.S. Department of Energy (DOE), Office of Science under contract DE-AC02-05CH11231, partially through Quantum Information Science Enabled Discovery (QuantISED) for High Energy Physics (KA2401032). Additional support is acknowledged from the U.S. Department of Energy, Office of Science, National Quantum Information Science Research Centers, Quantum Systems Accelerator. 

\bibliography{main}
\end{document}